\documentstyle[12pt]{article}
\begin{document}
\pagestyle{empty}
\hfill{ITP-SB-94-67}
\vskip 2cm
\centerline{\Large \bf Nonperturbative Corrections to the}
\centerline{\Large \bf Dilepton Cross Section}
\vskip .75cm
\centerline{Lyndon Alvero}
\vskip .5cm
\centerline{\it Institute for Theoretical Physics}
\centerline{\it State University of New York at Stony Brook}
\centerline{\it Stony Brook, NY 11794-3840}
\vskip 1.5cm
\centerline{December 1994}
\vskip 1cm
\centerline{\bf Abstract}

We study the sensitivity of the dilepton production cross section
to higher twist terms.  A method for calculating the resummed cross section is
introduced in moment space, where the leading higher twist terms can be
easily parametrized and calculated.  We find that a $1/Q$ power
correction may significantly affect the cross section at fixed-target energies.

\newpage
\section{Introduction}
\pagestyle{plain}
Large perturbative corrections are common in hadron-hadron scattering
cross sections in QCD \cite{ref:one}.  In this paper, we will
discuss corrections associated with soft gluons that produce large
threshold effects.  In the case of dilepton production cross sections, such
large corrections arise when most of the partonic momenta is carried off by
the lepton pair.  We will exhibit a class of related nonperturbative
corrections that can also have a significant impact on this cross section.

Several methods for resumming threshold corrections have been developed
\cite{ref:resum}.  Recently, resummed cross sections for dilepton production
at fixed-target energies have been calculated \cite{lahc1,lahc2} in principal
value resummation (PVR) \cite{hcgs1,hcgs2}.  In this scheme, the large
corrections exponentiate into leading and next-to-leading exponents, $E_L$
and $E_{NL}$, respectively.  These exponents contain the effects of the
one-($E_L$) and two-loop ($E_{NL}$) anomalous dimensions and
running coupling, and are each defined by a principal value integration.
Moreover, because singularities associated with
the infrared (IR) behavior of the QCD running coupling are avoided, no explicit
IR cutoffs need to be imposed in principal value resummation.
An analysis of perturbative resummation \cite{hcgs2,gkgs} shows the presence
of nonperturbative power suppressed (higher twist) contributions beginning at
${\cal O}(1/Q)$.

It was found in \cite{lahc2} that for protons on fixed-targets, the resummed
hard part (in PVR) when combined with parton distributions from global fits
led to cross sections that overestimate the data by roughly 30 to 50\%.  It was
also observed in \cite{lahc2} that the discrepancy could be due to higher twist
effects.  The aim of this paper is to explore this possibility.

In order to analyze the sensitivity of principal value resummation
to higher twist effects, we introduce a technique for calculating cross
sections which exploits the exponentiation of large corrections
in moment-space.  We shall show below that to a good approximation, the
partonic flux may be approximated by a polynomial in $z=Q^2/\hat{s}$, with $Q$
the pair mass and $\hat{s}$ the partonic center-of-mass invariant mass squared.
In this method, the resummed cross section becomes
a simple and finite sum of exponentials with coefficients that are functions
of energy alone, and most importantly, in which the higher twist effects are
easily parametrized and calculated.

The paper is organized as follows:  in section 2, we describe in detail how
we calculate the resummed cross section in moment space.  Here, we will focus
only on mass-rapidity distributions at zero rapidity, although our technique
can be applied to other distributions as well.  In section 3, we present
numerical results and compare with experiment.  We will find that the
resulting resummed cross section in moment space, combined with terms
proportional to $M/Q$, is sensitive to $M$ of the order of hundreds of MeV.
Indeed, we find good agreement with data when $M\sim -1$ GeV, for which the
higher twist contribution effectively cancels the effects of perturbative
resummation.
This, in effect, confirms that the cross section may be
quite sensitive to higher twist, at least for fixed-target energies.
Finally, in section 4, we give
a summary and conclusions. Some numerical details are discussed in an appendix.

\section{The cross section in moment-space}
Schematically, the dilepton-production reaction is given by
\begin{equation}
h_1(p_1)+h_2(p_2)\to l{\bar l}(Q^\mu)+X,
\label{drellyan}
\end{equation}
where $Q^2$ is the lepton-pair mass squared.  We will use the standard notation
$s=(p_1+p_2)^2$ and $\tau=Q^2/s$.

At fixed-target energies, where the $Z$ boson may be neglected,
principal value resummation provides the following
expression (in the DIS scheme), for the differential cross section
in dilepton production at zero rapidity, \cite{lahc2}
\begin{equation}
{d^2\sigma \over dQ^2dy}\biggl|_{y=0}\simeq \sigma_B(Q^2)\sum_f e_f^2
\int_{\tau}^1 dz\,  \omega_{f\bar f}(z,\alpha){{\cal F}_{f\bar f}^{y=0}(\tau/z)
\over z},
\label{dQdymain}
\end{equation}
with
\begin{equation}
\sigma_B(Q^2)={4\pi\alpha_e^2 \over 9Q^2s},
\label{sigB}
\end{equation}
\begin{equation}
\omega(z,\alpha)=A(\alpha)\biggl(\delta (1-z)-\biggl[e^{E({1\over 1-z},
\alpha)}\, {\sin [\pi P_1({1\over 1-z},\alpha)]\Gamma(1+P_1({1\over
1-z},\alpha))\over \pi(1-z)}\biggr]_+\biggr),
\label{omegres}
\end{equation}
\begin{equation}
{\cal F}_{f\bar f}^{y=0}(\tau/z)=F_{f/h_1}(\sqrt {\tau/z},Q)F_{\bar
f/h_2}(\sqrt {\tau/z},Q).
\label{partlum}
\end{equation}
The sum in eq.~(\ref{dQdymain}) is over all active quark and anti-quark
flavors, $\alpha_e$ is the (electromagnetic) fine structure constant,
$\alpha$ is proportional to the QCD running coupling (see eq.~(\ref{defat})
below) and $F_{f/h}(x,Q)$ denotes the
parton distributions.  The function $A(\alpha)$, which involves the resummation
of large, $z$-independent Sudakov terms, is given explicitly in appendix A,
while we define
\begin{equation}
P_1(x,\alpha)={d\over dx}E(x,\alpha).
\label{defP1}
\end{equation}
The function $E(z,\alpha)$, which exponentiates the large corrections, is,
as mentioned above, given as a sum of leading and nonleading contributions.
In moment-space, it takes the form
\begin{equation}
E(n,\alpha)=E(n,\alpha)_L+E(n,\alpha)_{NL}.
\label{defEn}
\end{equation}
The leading exponent is given explicitly by \cite{hcgs2}
\begin{equation}
E(n,\alpha)_L=\alpha(g_1^{(1)}I_1-g_2^{(1)}I_2),
\label{Elead}
\end{equation}
with
\begin{eqnarray}
I_1(n,t)& \equiv & 2I(n,t/2)-I(n,t),\nonumber \\
I(n,t)&=&t\int_Pd\zeta\biggl({\zeta^{n-1}-1\over 1-\zeta}\biggr)
\ln(1+(1/t)\ln(1-\zeta)),\nonumber \\
I_2(n,t)& \equiv & \int_Pd\zeta\biggl({\zeta^{n-1}-1\over 1-\zeta}\biggr)
{1\over 1+(1/t)\ln(1-\zeta)},
\label{IsElead}
\end{eqnarray}
where
\begin{equation}
\alpha\equiv \alpha_s(Q^2)/\pi, \, t\equiv 1/(\alpha b_2),
\label{defat}
\end{equation}
with
\begin{equation}
b_2=(11C_A-2n_f)/12.
\label{defb2}
\end{equation}
It will be sufficient to consider only the leading exponent $E(n,\alpha)_L$ in
the following discussion.  For completeness, however, we have reproduced the
explicit form of the next-to-leading exponent $E(n,\alpha)_{NL}$, as well as
the constants $g_1^{(1)}, g_2^{(1)}$ (eq.~(\ref{Elead})) in appendix A.

We are now ready to discuss the higher twist corrections to the resummed
cross section $d^2\sigma/dQ^2dy|_{y=0}$ (eq.~(\ref{dQdymain})).  To do so, it
will be necessary to go to moment- or $n$-space, where the
resummed hard part of eq.~(\ref{omegres}) takes the simple form
\begin{eqnarray}
\tilde{\omega}_{f\bar f}(n,\alpha)&\equiv&\int_0^1dz z^{n-1}\omega_{f\bar f}
(z,\alpha)\nonumber \\
&=&A(\alpha)e^{E(n,\alpha)},
\label{omegmom}
\end{eqnarray}
with $E(n,\alpha)$ given by eqs.~(\ref{defEn}-\ref{IsElead}) and $A(\alpha)$
by eq.~(\ref{fcnAexa}) of appendix A.

The leading higher twist term implied by the resummed exponent $E(n,\alpha)$
is of the form \cite{hcgs2,gkgs}
\begin{equation}
n{\Lambda \over Q}.
\label{nhitwist}
\end{equation}
This can be derived in the following way \cite{hcgs2}.
Consider the leading exponent
$E(n,\alpha)_L$, which depends on $I_1$ and $I_2$ (eq.~(\ref{IsElead})).  By a
change of variables, $\ln(1-\zeta)\rightarrow x'$, and using
\begin{equation}
(1-r)^{n-1}-1=\sum_{m=1}^{\infty}{(1-n)_m\over m!}r^m,
\label{intmed}
\end{equation}
where $(a)_m\equiv \Gamma(a+m)/\Gamma(a)$ is the Pochammer symbol, $I(n,t)$
in eq.~(\ref{IsElead}) can be rewritten as
\begin{equation}
I(n,t)=t\sum_{m=1}^{\infty}{(1-n)_m\over m!}{\cal P}\int_{-\infty}^0dx'
e^{mx'}\ln(1+x'/t),
\label{preInt}
\end{equation}
where ${\cal P}$ denotes a principal value integration.  After several more
simple steps, one obtains
\begin{equation}
I(n,t)=-\sum_{m=1}^{\infty}{(1-n)_m\over m!m^2}{\cal E}(mt),
\label{fInt}
\end{equation}
where the function ${\cal E}$ is proportional to the exponential integral
function:
\begin{equation}
{\cal E}(x)\equiv xe^{-x}{\cal P}\int_{-\infty}^xdy{e^y\over y}.
\label{exponint}
\end{equation}
Similarly, it can easily be shown that \cite{hcgs2}
\begin{equation}
I_2(n,t)=\sum_{m=1}^{\infty}{(1-n)_m\over m!m}{\cal E}(mt).
\label{fInt2}
\end{equation}

The leading higher twist term of $E(n,\alpha)$ comes from the
$2\alpha g_1^{(1)}I(n,t/2)$ term that appears in eq.~(\ref{Elead}).  Taking
only the leading $(m=1)$ term in the sum in eq.~(\ref{fInt}), we get
\begin{equation}
2\alpha g_1^{(1)}I(n,t/2)\sim \biggl(n{\Lambda \over Q}\biggr){\cal P}
\int_{-\infty}^{t/2}dy{e^y\over y},
\label{ltwist}
\end{equation}
where we have used the 1-loop form of $\alpha$, so that $t$ (eq.~(\ref{defat}))
is simply, $t=\ln Q^2/\Lambda^2$.  Since the pole of the integrand is at $y=0$,
independent of $Q$, the basic structure of the leading power correction
to $E(n,\alpha)$ is $1/Q$.  This argument is specific to principal value
resummation, but the result is quite general, as discussed in \cite{gkgs},
where it was also shown that this power correction is actually
induced by the leading infrared renormalon \cite{ref:IRren}.

We now digress on the issue of including higher twist terms in momentum- or
$z$- space.  With the correspondence \cite{ref:korch}
\begin{equation}
n\leftrightarrow {1 \over 1-z},
\label{nzcorr}
\end{equation}
the leading power correction can be parametrized by making the substitution
\begin{equation}
E(z,\alpha)\rightarrow E(z,\alpha)-{c \over 1-z} {\Lambda \over Q},
\label{expwhit}
\end{equation}
for some constant $c$, in eq.~(\ref{omegres}).  We then find that the direct
calculation of the resummed cross section given by eq.~(\ref{dQdymain}) is
hopeless, because an essential singularity in
$\omega_{f\bar f}(z,\alpha)$ (eq.~(\ref{omegres})) will be generated at $z=1$.
This highlights one of the advantages of working in $n$-space, for,  as we
shall now see, the higher twist term (eq.~(\ref{nhitwist})) is in a form that
makes it easy to calculate its effects in the resummed cross section.

To write the cross section in terms of moments, we first need to express
the sum involving the parton luminosity as a polynomial in $z$:
\begin{equation}
\sum_f e_f^2{\cal F}_{f\bar f}^{y=0}(\tau/z)\equiv\sum_{n=1}^{n_{max}}
C_n(\tau)z^n + \Delta{\cal F}(\tau,z),
\label{polydef}
\end{equation}
where $\Delta{\cal F}$ denotes the error in the polynomial
approximation.  Some details on how $n_{max}$ is determined, as well as how
the interpolating polynomial is obtained, are described in appendix B.

Using eq.~(\ref{polydef}) in eq.~(\ref{dQdymain}), we obtain
\begin{equation}
{d^2\sigma \over dQ^2dy}\biggl|_{y=0}\simeq \sigma_B(Q^2)\sum_{n=1}^{n_{max}}
C_n(\tau)\int_{\tau}^1dz z^{n-1}\omega_{f\bar f}(z,\alpha)+\sigma_B(Q^2)
\int_{\tau}^1dz\, \omega_{f\bar f}(z,\alpha){\Delta{\cal F}(\tau,z) \over z}.
\label{ninterm}
\end{equation}

Combining eqs.~(\ref{omegmom}) and (\ref{ninterm}), and applying several
straighforward manipulations, we find
\begin{equation}
{d^2\sigma \over dQ^2dy}\biggl|_{y=0}\simeq \sigma_B(Q^2)A(\alpha)
\sum_{n=1}^{n_{max}}C_n(\tau)e^{E(n,\alpha)}+CT_1+CT_2.
\label{omegmom2}
\end{equation}
The explicit forms of the correction terms $CT_1$ and $CT_2$ will be given
shortly.  To study the sensitivity of eq.~(\ref{omegmom2}) to higher twist, we
insert a term proportional to $n\Lambda/Q$ in the exponent, so that now the
$n$-space resummed cross section is given by
\begin{equation}
{d^2\sigma \over dQ^2dy}\biggl|_{y=0}\simeq \sigma_B(Q^2)A(\alpha)
\sum_{n=1}^{n_{max}}C_n(\tau)\exp\biggl[E(n,\alpha)-{cn\Lambda \over Q}
\biggr]+CT_1+CT_2,
\label{ommouse}
\end{equation}
where $\Lambda$ is the QCD scale, $c$ is a constant, and the correction terms
are
\begin{eqnarray}
CT_1&=&\sigma_B(Q^2)\int_{\tau}^1dz\, \omega_{f\bar f}(z,\alpha){\Delta{\cal F}
(\tau,z) \over z},\nonumber \\
CT_2&=&\sigma_B(Q^2)A(\alpha)\sum_{n=1}^{n_{max}}C_n(\tau)\int_0^{\tau}dz
z^{n-1}e^{E(z,\alpha)}\, {\sin [\pi P_1(z,\alpha)]\Gamma(1+
P_1(z,\alpha)) \over \pi(1-z)},
\label{corterms}
\end{eqnarray}
with\footnote{Note that for brevity, we have represented the $1/(1-z)$
dependence of the functions in $CT_2$ as simply $z$.}
$\omega_{f\bar f}(z,\alpha)$ given by eq.~(\ref{omegres}), and
\begin{equation}
{\Delta{\cal F}(\tau,z) \over z}=\sum_fe_f^2{{\cal F}_{f\bar f}^{y=0}(\tau/z)
\over z}-\sum_{n=1}^{n_{max}}C_n(\tau)z^{n-1}.
\label{delF}
\end{equation}

We will use eq.~(\ref{ommouse}) to calculate resummed cross sections in the
next section.  Note that in this form, the cross section is simply a finite sum
of exponentials weighted by the coefficients of the interpolating polynomial,
up to corrections ($CT_1$ and $CT_2$) which are expected to be very small,
since $\Delta{\cal F}(\tau,z)$ is small.

\section{Numerical results}

For our numerical calculations, we chose to work with the kinematics of the
E605 experiment \cite{e605}, which studied proton beam interactions with a
copper target at ${\sqrt s}=$38.8 GeV.  We have used the CTEQ2D
distributions with $\Lambda$ at 4(5) flavors equal to 0.235(0.155) GeV.
We have also included all finite 1-loop contributions (see \cite{lahc2}) in all
the resummed cross sections that we are now going to present.  In the
discussions to follow, we will simply denote the cross section
$d^2\sigma/dQ^2dy|_{y=0}$ by $\sigma$.  Hence, $\sigma(n_{max},c\Lambda)$ will
refer to the cross sections computed in $n$-space from eq.~(\ref{ommouse})
as described above, for a given $n_{max}$ and $c\Lambda$,
and $\sigma_z$ the cross section computed by the usual method as an
integral over $z$ (eq.~(\ref{dQdymain})) with no explicit higher twist,
at the same value of $Q$.  We also point out that the cross section defined by
eq.~(\ref{ommouse}) is dominated by the sum over $n$, with the correction terms
giving a contribution of not more than 1.6\%.

To determine $n_{max}$, we used the method described in \cite{ref:linreg}.  The
basic idea is that by adding a $z^{n+1}$ term to a degree-$n$ polynomial fit
to a particular set of data (the parton luminosity (eqs.~(\ref{partlum},
\ref{polydef}))), the deviation of the interpolation from the data
is decreased.  However, for some degree $n_{max}$, the addition of higher
powers of $z$ no longer decreases the deviations by a statistically significant
amount.  We have found that for the data points with which we compare,
$n_{max}$ ranges from 13 to 16 only.  We also mention that, although the choice
of $z_{min}$, which defines
the range of validity of the interpolation (see appendix B), somewhat
affects the value of $n_{max}$, we have verified that the $n$-space
resummed cross section is insensitive to $z_{min}$.  That is,
$\sigma(n_{max},c\Lambda)\simeq \sigma(n_{max}',c\Lambda)$, where
$n_{max} (n_{max}')$ is obtained when $z_{min} (z_{min}')$ is used in the
interpolation.

We plot in fig.\ 1 the resummed cross sections calculated in moment-space and
scaled relative to the corresponding $z$-space cross sections taken from
the resummed curve (solid line in fig.\ 2).  The points denoted by circles,
triangles and squares correspond to $n$-space cross sections calculated using
eq.~(\ref{ommouse}) with $c\Lambda$ equal to 0, 0.5 and 1 GeV, respectively
[recall that we take $c=0$ for $\sigma_z$].

We see that the $n$-space resummed cross section $\sigma(n_{max},0)$ is
different from the $z$-space cross section $\sigma_z$.  The two differ,
however, by no more than 12\%, with $\sigma(n_{max},0)$ initially smaller but
eventually becoming larger than $\sigma_z$.  This difference already
illustrates the range of sensitivity to higher twist effects that are implicit
in any resummation prescription.  We can also infer from fig.\ 1 that for
$c\Lambda=1 (0.5)$ GeV, the higher twist effects are about 27 to 23\%
(14 to 11\%) of the resummed cross section $\sigma(n_{max},0)$, decreasing
with $\tau$.

In fig.\ 2, we plot the data of experiment E605 and reproduce the curves of
fig.\ 9d of \cite{lahc2} (dashed curve=2-loop
and solid line=resummed cross section).  For comparison, we also include the
results obtained for $\sigma(n_{max},1$GeV) (dotted curve).  We see that, with
this value of $c\Lambda$, the $n$-space resummed cross section gives quite a
good fit to the data for this range of $\tau$.  In fact, the
$\sigma(n_{max},1$GeV) curve is almost identical to the 1-loop curve in
fig.\ 9d of \cite{lahc2}.

\section{Discussions and conclusions}

In this paper, we have derived an expression in moment-space for
the cross section in principal value resummation.  In this form, the
cross section becomes a finite sum of exponentials, and higher twist terms
can be included in an easy, straightforward manner.  This is in contrast
with the resummation formula written in momentum- or $z$-space, where it is not
possible to treat the higher twist in an analogous fashion.

Direct comparison with experiment shows that a $1/Q$ correction,
with a coefficient of order 1 GeV, in this variant form of principal value
resummation, agrees with experiment, while the purely perturbative resummed
cross sections in \cite{lahc2} overestimate the data.
It seems that such a higher twist term is large enough to cancel the effects of
resummation, bringing down the resummed cross section to the size of the
1-loop cross section.
We also note that this size of the higher twist is comparable to those found
in $e^+e^-$ event shape variables \cite{webber}.

We also mention that for the particular distribution we are concerned with,
$d^2\sigma/dQdy|_{y=0}$, it was found in \cite{ref:prwvn} that already at
${\cal O}(\alpha_s)$, the non-singular terms in the hard part account for
about 20\% of the total cross section.  Although we have included all such
1-loop finite contributions to our resummed cross sections, there is still
some uncertainty due to similar non-singular contributions at higher orders.
However, since we are mostly concerned with the structure and effects of the
leading higher twist term, and not the precise calculation of cross sections,
our conclusions are not affected by this uncertainty.

It should be noted that we have not really explained why the higher twist
coefficient should be as big as 1 GeV.  It could be that it is actually
smaller, in which case, the resummed cross section will still overestimate the
data but less than the amount found in \cite{lahc2}.  The remaining discrepancy
could then be explained by the other factors alluded to in \cite{lahc2}:
the importance of higher order effects when combining deeply inelastic
scattering (DIS) and hadron-hadron data in global fits, the theoretical
uncertainty in $A(\alpha)$, and the effects of finite higher order terms in
the hard part.  Probably, the full cross section comes from a combination of
all such effects.  Here, however, we have shown that higher twist does play
a role.

\vskip 1in
\centerline{ACKNOWLEDGMENTS}

I would like to thank George Sterman for many valuable discussions, and
for his constant support and guidance.  I would also like to thank W.\ L.\
van Neerven for a helpful discussion.  This work is supported in part by the
National Science Foundation, under grant PHY 9309888.

\newpage

\appendix{\bf Appendix A The functions $E(n,\alpha)_{NL}$ and $A(\alpha)$}

Here, we write down the explicit expressions for the functions $E(n,
\alpha)_{NL}$ and $A(\alpha)$ which were described in section 2.  The
next-to-leading exponent $E(n,\alpha)_{NL}$ is given by \cite{lahc1}
\begin{equation}
E(n,\alpha)_{NL}=
\alpha(g_1^{(1)}J_1-g_2^{(1)}J_2)+\alpha^2(g_1^{(2)}K_1-g_2^{(2)}K_2),
\label{Enlead}
\end{equation}
with
\begin{eqnarray}
J_1& \equiv & (\alpha b_3/b_2)\int_P d\zeta\biggl({\zeta^{n-1}-1\over 1-\zeta}
\biggr)
\int_0^\zeta {dy\over 1-y}{\ln(1+(1/t)\ln[(1-\zeta)(1-y)])\over (1+(1/t)
\ln[(1-\zeta
)(1-y)])^2},\nonumber \\
J_2& \equiv & -(\alpha b_3/b_2)\int_P d\zeta\biggl({\zeta^{n-1}-1\over 1-\zeta}
\biggr)
{\ln(1+(1/t)\ln(1-\zeta))\over (1+(1/t)\ln(1-\zeta))^2},\nonumber \\
K_1& \equiv & -\int_Pd\zeta\biggl({\zeta^{n-1}-1\over 1-\zeta}\biggr)
\int_0^\zeta{dy
\over 1-y}{1\over (1+(1/t)\ln[(1-\zeta)(1-y)])^2},\nonumber \\
K_2& \equiv & \int_Pd\zeta\biggl({\zeta^{n-1}-1\over 1-\zeta}\biggr)
{1\over (1+(1/t)\ln(1-\zeta))^2},
\label{JKsnlead}
\end{eqnarray}
where $\alpha$ and $t$ are defined in eq.~(\ref{defat}).
All integrals are evaluated over a principal value contour as defined in
\cite{hcgs2,lahc1}.  The various constants are given by\footnote{The parameter
$g_2^{(2)}$ can be found from Table 1 of \cite{ref:chdav}, where $B^{(2)}=4
g_2^{(2)}$.  The relation between the $g_j^{(i)}$ and $B^k$ is described in
\cite{hcgs1}.  The extra factor of $1/4$ in eq.~(\ref{constants}) is due to our
expansion in $\alpha\equiv \alpha_s/\pi$ rather than $\alpha_s/2\pi$ as
in \cite{ref:chdav}.}
\begin{eqnarray}
&g_1^{(1)}=2C_F,\ g_2^{(1)}=-{3\over 2}C_F,\ g_1^{(2)}=C_F\left[C_A
\left({67\over 18}-{\pi^2\over 6}\right)-{5n_f\over 9}\right],\nonumber \\
&g_2^{(2)}={1\over 4}\biggl[C_F^2(\pi^2-{3\over 4}-12\zeta(3))+C_AC_F({11
\over 9}\pi^2-{193\over 12}+6\zeta(3))+{C_F\over 2}(-{4\over 9}\pi^2+{17
\over 3})\biggr],\nonumber \\
&b_3=(34C_A^2-(10C_A+6C_F)n_f)/48,
\label{constants}
\end{eqnarray}
where $n_f$ is the number of flavors, and $C_A, C_F$ are color factors.  The
parameter $b_2$ is defined by eq.~(\ref{defb2}).

We also point out here that in the calculation of $E_{NL}$ for this paper,
the term proportional to the integral $K_2$ was not included.  This term,
strictly speaking, yields logarithms beyond nonleading order, as defined in
\cite{hcgs1,hcgs2}.  This treatment differs from the calculations in
\cite{lahc1,lahc2}, where the same term was kept.  However, we have checked
that the contribution from the above $K_2$ term is no more than 1\% of the
total cross section.  In particular, the resummed curve in fig.\ 2,
which excludes the contributions from the $K_2$ term, is nearly identical to
the solid curve of fig.\ 9d in \cite{lahc2}.

The explicit form of $A(\alpha)$, to the same order, is \cite{lahc2}
\begin{eqnarray}
A(\alpha)&=&\biggl(1+2C_F\alpha+b\alpha^2\biggr)\exp\Biggl(
{\alpha \over 2}C_F\biggl[{\pi^2 \over 3}-3\biggr]+{G^{(1)} \over b_2}\ln r
+\pi G^{(2)}{\sin \theta \over r}\alpha^2\nonumber \\
& &+{\gamma_K^{(1)} \over 2 b_2^2}{1 \over \alpha}(\pi b_2
\theta  \alpha-\ln r)+{\gamma_K^{(2)} \over 2 b_2^2} \ln r\Biggr),
\label{fcnAexa}
\end{eqnarray}
with
\begin{eqnarray}
r&=&\biggl[1+(\pi b_2\alpha)^2\biggr]^{1 \over 2},\ \theta=\arctan(\pi b_2
\alpha),\ G^{(1)}={3 \over 2}C_F,\nonumber \\
G^{(2)}&=&C_F^2\biggl[{3 \over 16}-{7 \over 3}\zeta(3)+{23
\over 6}\zeta(2)\biggr]+C_AC_F\biggl[{2545 \over 432}+{11 \over 12}
\zeta(2)-{13 \over 4}\zeta(3)\biggr]\nonumber \\
& & +n_fC_F\biggl[-{209 \over 216}-{1 \over 6}\zeta(2)\biggr],\nonumber \\
\gamma_K^{(1)}&=& 2C_F,\,
\gamma_K^{(2)}= C_AC_F\biggl[{67 \over 18}-\zeta(2)\biggr]-{5 \over 9}n_fC_F,
\end{eqnarray}
and
\begin{eqnarray}
b&=& C_F^2\biggl[{-23 \over 720}\pi^4-{35 \over 96}\pi^2
+{15 \over 2}\zeta(3)+{15 \over 8}\biggr]
+C_AC_F\biggl[{215 \over 144}+{175 \over 216}\pi^2
-{49 \over 12}\zeta(3)-{17 \over 1440}\pi^4\biggr]\nonumber \\
& &+n_fC_F\biggl[{1 \over 3}\zeta(3)-{19 \over 72}-{7 \over 54}\pi^2\biggr],
\label{moreconst}
\end{eqnarray}
where $\zeta$(s) is the Riemann Zeta function.

\vskip 1in

\appendix{\bf Appendix B Interpolating Polynomials}

The interpolating polynomial that we require is defined by eq.~(\ref{polydef}),
repeated below:
\begin{eqnarray}
\sum_f e_f^2{\cal F}_{f\bar f}^{y=0}(\tau/z)\equiv\sum_{n=1}^{n_{max}}
C_n(\tau)z^n + \Delta{\cal F}(\tau,z).\nonumber
\end{eqnarray}
For a particular value of $Q$, we generate a set of values of the left-hand
side of the above equation in a certain range of $z$, $z_{min}<z<1$.
The parameter $z_{min}$ is selected such that ${\cal F}_{f\bar f}^{y=0}
(\tau/z)\sim 0$ for $z<z_{min}$.  The typical values of $z_{min}$ that we used
vary from 0.075 to 0.2.
We then use the Fortran 77 IMSL package on each set of values to obtain the
interpolating polynomial, or equivalently, the coefficients $C_n(\tau)$.
The polynomials that we obtain fit the luminosity data very well, with
percentage differences of the order of $10^{-5}$ or better.

In principle, given a set of data points, one can always fit a polynomial of
degree $n_{max}$, up to $n_{max}=$ number of points.  However, as described in
section 3, a method exists for determining a non-trivial $n_{max}$.
One expects that the polynomial can fit the luminosity data very well in the
region $z>z_{min}$.  Outside this fit region, care must be taken that the
selected
polynomial does not behave wildly.  To be specific, the lowest value
that $z$ (or $z_{min})$ can take is $\tau$, where the luminosity vanishes.
It must then be checked that for $z<z_{min}$, the polynomial
fits are free of large oscillations.  We have verified that the fits we
have obtained are indeed well-behaved in this region of $z$.

\newpage

\appendix{\bf FIGURE CAPTIONS}
\begin{description}
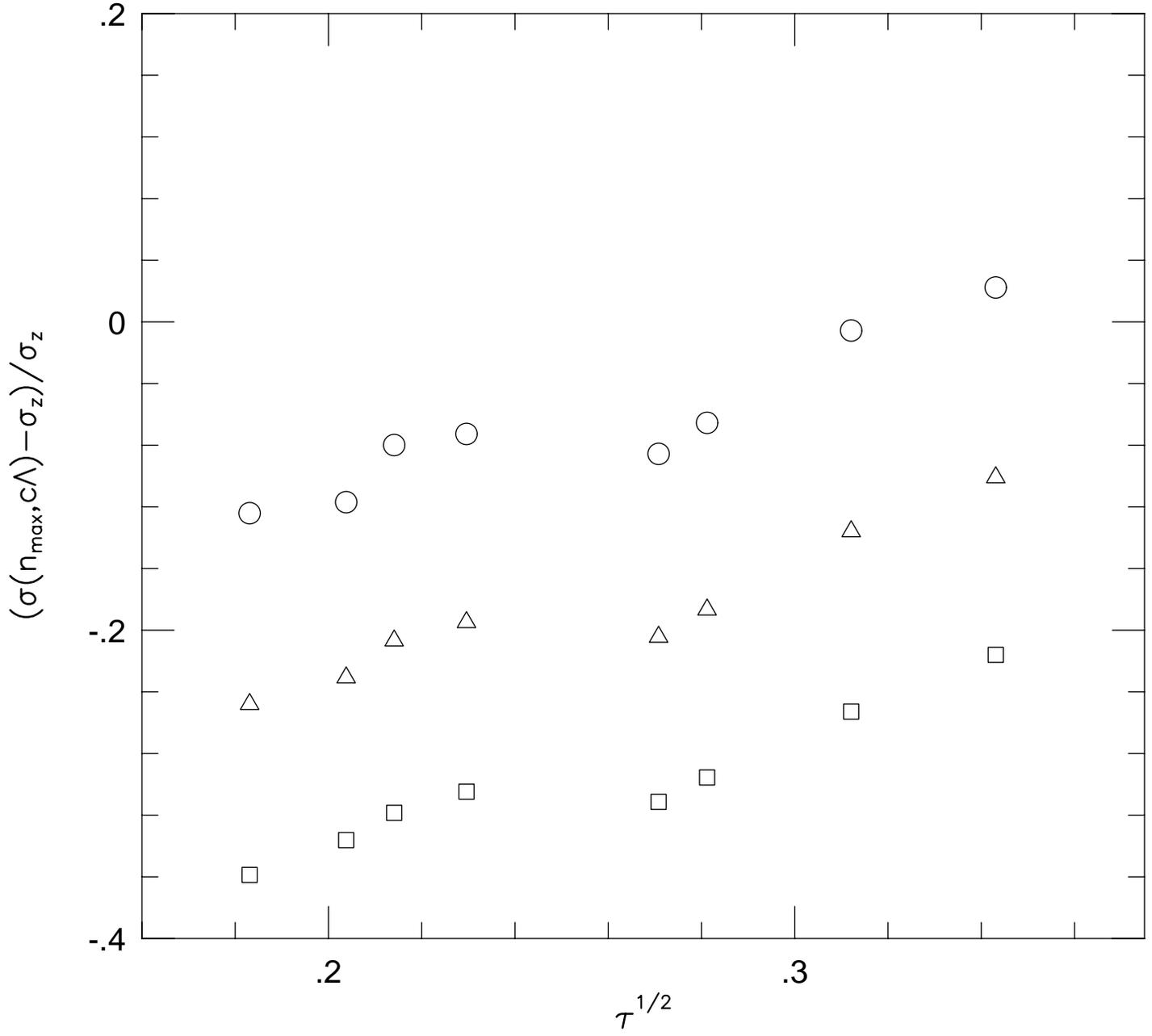
\item{Figure 1.} Fractional deviations of the resummed cross section in
$n$-space,\\ $\sigma(n_{max},c\Lambda)$ (eq.~(\ref{ommouse})), from the purely
perturbative cross section, $\sigma_z$ (eq.~(\ref{dQdymain})).\\
$c\Lambda=0$ (circles); $c\Lambda=0.5$ GeV (triangles);
$c\Lambda=1$ GeV (squares).
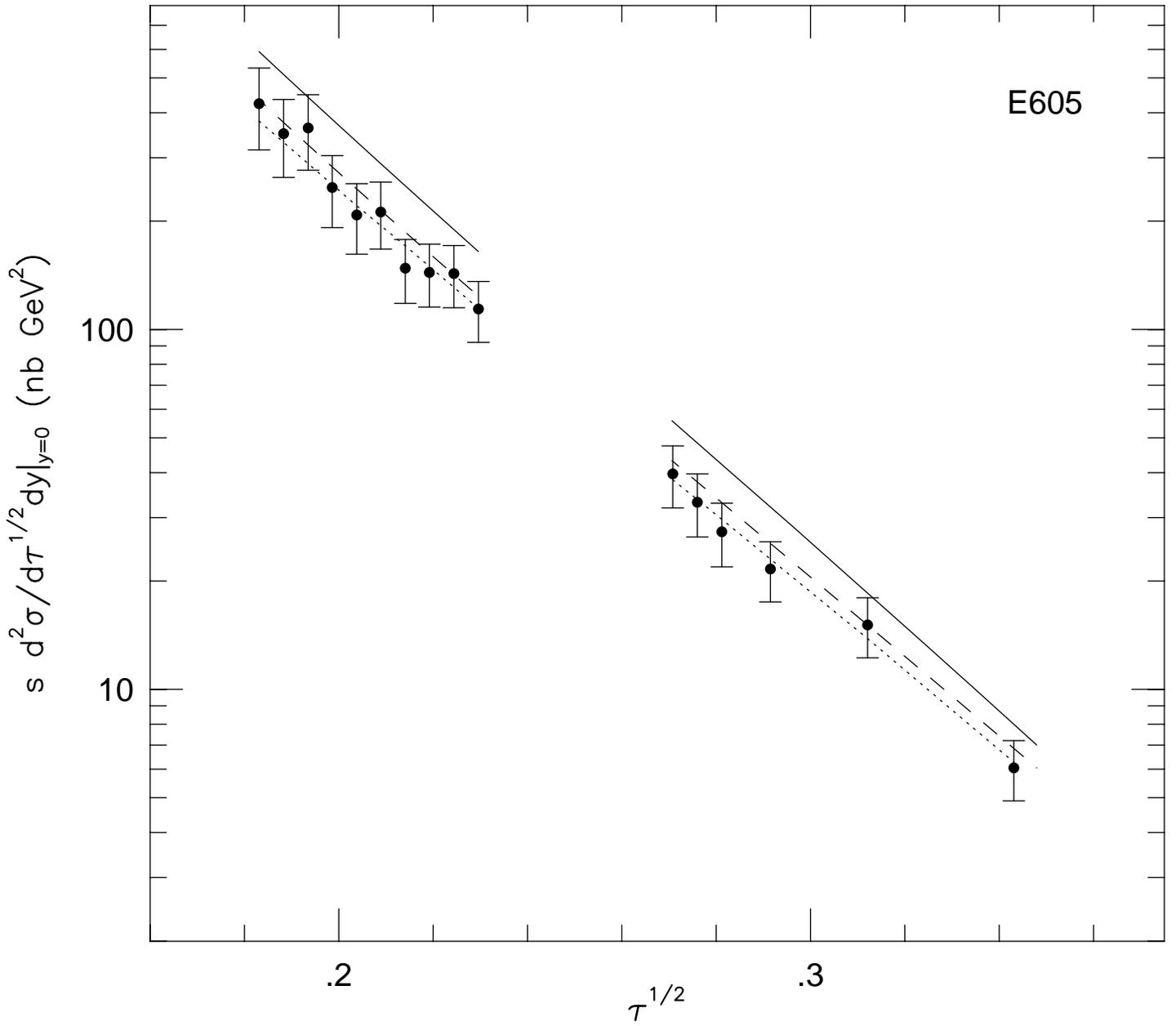
\item{Figure 2.} Comparison with E605 data.\\
Solid=resummed $\sigma_z$; Dashed=2-loop; Dotted=resummed
$\sigma(n_{max},c\Lambda=1$ GeV).

\end{description}

\end{document}